\documentclass[11pt]{article}
\usepackage{amsmath}
\usepackage{amsfonts}
\usepackage{graphicx}
 \usepackage{amsthm}
 \usepackage{pictexwd,dcpic}

\newtheorem{mythm}{Theorem}

\begin{document}

\title{Three player, Two Strategy, Maximally Entangled Quantum Games}
\author{Aden Ahmed\footnote{PhD candidate, Department of Mathematics and Statistics, Portland State University, Portland, Oregon USA. Email: ahmedao@pdx.edu.} \hspace{.1in} Steve Bleiler\footnote{Professor of Mathematics, Department of Mathematics and Statistics, Portland State University. Email: bleilers@pdx.edu.} \hspace{.1in}Faisal Shah Khan\footnote{PhD candidate, Department of Mathematics and Statistics, Portland State University. Adjunct Faculty, University of Portland, Portland Oregon, USA. Email: faisal@pdx.edu.}}
\maketitle

\begin{abstract}
We develop an octonionic representation of the payoff function for a three player, two strategy, maximally entangled quantum game. 
\end{abstract}

\section{A Formalism for Quantization}

Up until now, to quantize games most authors have, like Meyer, focused their efforts on the quantization of the players' strategy spaces, essentially the domain of the payoff function that defines the game to be quantized (see for example \cite{Eisert, Gutoski, Marinatto, Meyer}). The principal technique is to identify these strategy spaces with an orthogonal basis of some quantum system, in order that players may now take superpositions or even mixed superpositions of strategic choices by acting on the system via quantum operations. In addition, players may now correlate their strategic choices via the entanglement of the joint states of the system. Frequently, mere access to the higher randomization of superposition (as opposed to real probabilistic combination) or the correlation of strategic choices via entanglement allows payoffs to the players superior to those available in the game and its classical extensions. In \cite{Bleiler1} a mathematical formalism for game quantization that focuses on the quantization of the {\it payouts} of the original game $G$ to be quantized, and expresses the quantized version as a (proper) extension of the original payout function in the set-theoretic sense (just as in the classical case) is developed. We brielfy recall this formalism next.

Classically, during extensions, one constructs probability distributions over the outcomes of a game $G$. We now wish to pass to a more general notion of randomization, that of quantum superposition. Begin then with a Hilbert space $\mathcal{H}$ that is a complex vector space equipped with an inner product. For the purpose here let us assume that $\mathcal{H}$ is finite dimensional, and that we have a finite set $X$ which is in one-to-one correspondence with an orthogonal basis $\mathcal{B}$ of $\mathcal{H}$. 

By a {\it quantum superposition} of $X$ with respect to the basis $\mathcal{B}$ we mean a complex projective linear combination of elements of $X$; that is, a representative of an equivalence class of complex linear combinations where the equivalence between combinations is given by non-zero scalar multiplication. Quantum mechanics calls this scalar a {\it phase}. When the context is clear as to the basis to which the set $X$ is identified, denote the set of quantum superpositions for $X$ as $QS(X)$. Of course, it is also possible to define quantum superpositions for infinite sets, but for the purpose here, one need not be so general. What follows can be easily generalized to the infinite case. 

As the underlying space of complex linear combinations is a Hilbert space, we can assign a length to each linear combination and, up to phase, always represent a projective linear combination by a complex linear combination of length 1. This process is called {\it normalization} and is frequently useful. 

For each quantum superposition of $X$ we can obtain a probability distribution over $X$ by assigning to each component the ratio of the square of the length of its coefficient to the square of the length of the combination. For example, the probability distribution produced from $\alpha x + \beta y$ is just
$$
\frac{\left|\alpha\right|^2}{\left|\alpha\right|^2+\left|\beta\right|^2}x+\frac{\left|\beta\right|^2}{\left|\alpha\right|^2+\left|\beta\right|^2}y
$$
Call this function $QS(X) \rightarrow \Delta(X)$ a {\it quantum measurement with respect to $X$}, and note that geometrically quantum measurement is defined by projecting a normalized quantum superposition onto the various elements of the normalized basis $\mathcal{B}$. Denote this function by $q_X^{meas}$, or if the set $X$ is clear from the context, by $q^{meas}$.

Now given a finite $n$-player game $G$, suppose we have a collection $Q_1, \dots, Q_n$ of non-empty sets and a \emph{protocol}, that is, a function $\cal{Q}$ $:\prod Q_i \rightarrow QS(\rm{Im}G)$. Quantum measurement $q_{\rm{Im}G}^{meas}$ then gives a probability distribution over $\rm{Im}G$. Just as in the classical mixed strategy case we can then form a new game $G^{\cal Q}$ by applying the expectation operator. Call the game $G^{\cal Q}$ thus defined to be the {\it quantization of $G$ by the protocol \cal{Q}}. Call the $Q_i$'s sets of {\it pure quantum strategies} for $G^{\cal Q}$. Moreover, if there exist embeddings $e'_i:S_i \rightarrow Q_i$ such that $G^{\cal Q} \circ \prod e'_i=G$, call $G^{\cal Q}$ a {\it proper} quantization of $G$. If there exist embeddings $e''_i:\Delta(S_i) \rightarrow Q_i$ such that $G^{\cal Q} \circ \prod e''_i=G^{mix}$, call $G^{\cal Q}$ a {\it complete} quantization of $G$. These definitions are summed up in the following commutative diagram. Note for proper quantizations the original game is obtained by restricting the quantization to the image of $\prod e'_i$. For general extensions, the Game Theory literature refers to this as ``recovering'' the game $G$. 

\begin{figure}[h]
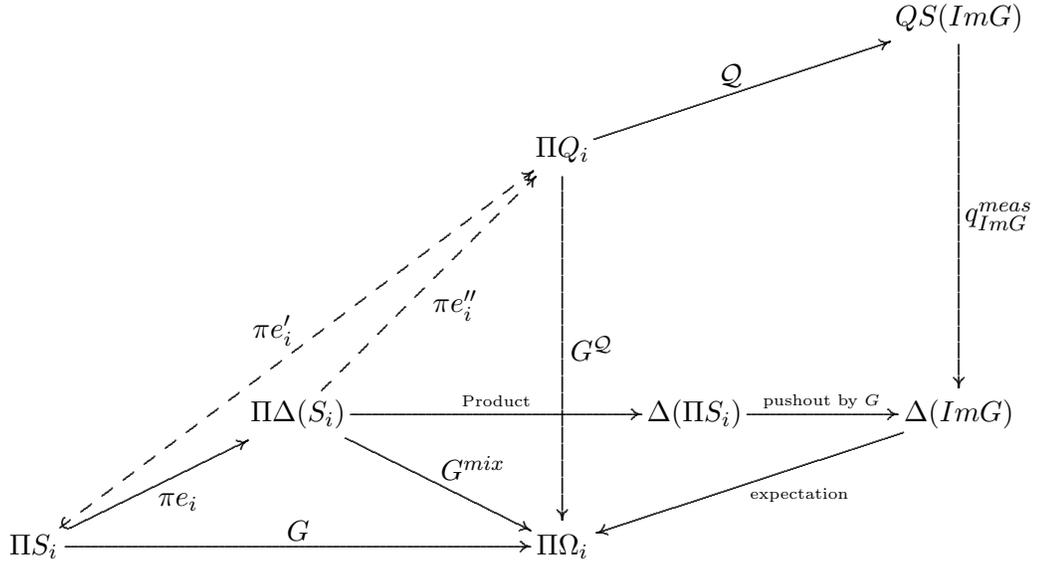

	\[ \begindc{0}[50]
\obj(4,3){$\Pi Q_i$}
\obj(7,4){$QS(ImG)$}
\obj(7,1){$\Delta(ImG)$}
\obj(5,1){$\Delta (\Pi S_i)$}
\obj(2,1){$\Pi \Delta(S_i)$}
\obj(4,0){$\Pi \Omega_i$}
\obj(0,0){$\Pi S_i$}
\mor{$\Pi Q_i$}{$QS(ImG)$}{$\cal Q$}
\mor{$\Pi Q_i$}{$\Pi \Omega_i$}{$G^{\cal Q}$}[\atleft, \solidarrow]
\mor{$QS(ImG)$}{$\Delta(ImG)$}{$q_{ImG}^{meas}$}
\mor{$\Delta(ImG)$}{$\Pi \Omega_i$}{$\mbox{\tiny{expectation}}$}
\mor{$\Pi S_i$}{$\Pi \Omega_i$}{$G$}
\mor{$\Pi \Delta(S_i)$}{$\Pi Q_i$}{$\pi e^{\prime \prime}_i$}[\atright, \dasharrow]
\mor{$\Pi S_i$}{$\Pi Q_i$}{$\pi e^{\prime}_i$}[\atleft, \dasharrow]
\mor{$\Pi S_i$}{$\Pi \Delta(S_i)$}{$\pi e_i$}[\atright, \injectionarrow]
\mor{$\Pi \Delta(S_i)$}{$\Delta (\Pi S_i)$}{$\mbox{\tiny{Product}}$}
\mor{$\Delta (\Pi S_i)$}{$\Delta(ImG)$}{$\mbox{\tiny{pushout by $G$}}$}
\mor{$\Pi \Delta(S_i)$}{$\Pi \Omega_i$}{$G^{mix}$}
\enddc \]
	\caption{A Quantization Formalism}
	\label{fig:ACompleteQuantizationOfG}
\end{figure}

Note that the definitions of $G^{mix}$ and $G^{\cal Q}$ show that a complete quantization is proper. Furthermore, note that finding a mathematically proper quantization of a game $G$ is now just a typical problem of extending a function. It is also worth noting here that nothing prohibits us from having a quantized game $G^{\cal Q}$ play the role of $G$ in the classical situation and by considering the probability distributions over the $Q_i$, creating a yet larger game $G^{m \cal{Q}}$, the {\it mixed quantization of G with respect to the protocol $\cal{Q}$}. For a proper quantization of $G$, $G^{m \cal{Q}}$ is an even larger extension of $G$. The game $G^{m \cal{Q}}$ is described in the commutative diagram of figure \ref{fig:ExtensionOfGByGQm}.

\begin{figure}[h]
\[ \begindc{0}[50]
\obj(0,2){$\Pi_{i=1}^n \Delta(Q_i)$}
\obj(2,2){$\Delta (\Pi_{i=1}^n Q_i)$}
\obj(5,2){$\Delta(ImG^{Q})$}
\obj(5,0){$\Pi_{i=1}^n \Omega_i$}
\obj(0,0){$\Pi_{i=1}^n Q_i$}
\mor{$\Delta (\Pi_{i=1}^n Q_i)$}{$\Delta(ImG^{Q})$}{$\mbox{\tiny{pushout by $G^Q$}}$}
\mor{$\Pi_{i=1}^n \Delta(Q_i)$}{$\Delta (\Pi_{i=1}^n Q_i)$}{$\mbox{\tiny{Product}}$}
\mor{$\Delta(ImG^{Q})$}{$\Pi_{i=1}^n \Omega_i$}{$\mbox{\tiny{expectation}}$}
\mor{$\Pi_{i=1}^n Q_i$}{$\Pi_{i=1}^n \Delta(Q_i)$}{$\pi \tilde{e}_i$}[\atright, \injectionarrow]
\mor{$\Pi_{i=1}^n Q_i$}{$\Pi_{i=1}^n \Omega_i$}{$G^{\cal Q}$}
\mor{$\Pi_{i=1}^n \Delta(Q_i)$}{$\Pi_{i=1}^n \Omega_i$}{$G^{m \cal{Q}}$}
\enddc \]	
	\caption{Extension of $G$ by $G^{m \cal{Q}}$}
	\label{fig:ExtensionOfGByGQm}
\end{figure}

Note that the quantum strategy sets $Q_i$ need not consist of quantum superpositions, although this is frequently the case. Indeed, protocols with classical inputs yielding quantum superpositions of the outcomes of certain games have already been posited \cite{Dahl, Iqbal}. These and some other specific protocols are discussed in the context of the formalism above in \cite{Bleiler1}.

As discussed in \cite{Bleiler1} and in part the following sections, the literature gives several protocols for quantizing one, two, and occasionally even multi-player games, some improper, some proper but not complete, and some yielding complete quantizations. Yet there is an ongoing debate in the literature as to what is the `correct' method of quantizing a game. The above formalism suggests that this is just the wrong question to ask, as under this formalism a given game can admit several different quantizations. It also makes clear that comparisons between various quantizations, between quantizations and various classical extensions, and between quantizations and the original game itself often amounts to comparing ``apples'' to ``oranges''.

\section{Mediated Quantum Communication via the EWL Protocol}

In classical mediated communication, players have a referee mediate their game and the communication of their strategic choices. For simplicity, assume our players have but two classical pure strategies to choose from. The communication of each players strategic choices is implemented by the sending of bits to the players, put into an initial state by the referee. Presumably players then send back their individual bits in the other state (Flipped) or in the original state (Un-Flipped) to indicate the choice of their second or first classical pure strategy respectively. The bits are then examined by the referee who then makes the appropriate payoffs. 

When the communication between the referee and the players is over quantum channels, Eisert, Wilkens and Lewenstein \cite{Eisert} have proposed families of quantization protocols that depend on the initial joint state prepared by the referee. Players and the referee communicate via {\it qubits}, a two pure state quantum system with a fixed observational basis.  In the EWL protocol the referee determines the payouts via a new observational basis corresponding to the actions of (No Flip, No Flip), (No Flip, Flip), (Flip, No Flip), (Flip, Flip) by the players. Players may choose from any physical operation (i.e. the Lie group $S(2)$) as pure quantum strategies (the $Q_i$'s in the formalism above) or even probabilistic combinations thereof (the $\Delta Q_i$'s in the formalism) for their strategic choices. The procedure above describes for each initial state $I$ a protocol ${\cal{Q}}_I$ and a quantized and mixed quantized game $G^{{\cal{Q}}_I}$ and $G^{{m\cal{Q}}_I}$ per the formalism.

If the initial state prepared by the referee is given in the Dirac notation by $\left|0\right\rangle \otimes \left|0\right\rangle$, then the EWL protocol is a complete quantization and is in fact equivalent to the classical game $G^{mix}$. But when the initial state is given by the maximally entangled state $\left|0\right\rangle \otimes \left|0\right\rangle+\left|1\right\rangle \otimes \left|1\right\rangle$, the EWL protocol remains complete and sets up an onto map from the product of the strategy spaces to $\Delta\left(ImG\right)$.

\section{Games with Two Strategic Choices}
We consider classical games where each player has exactly two pure strategic choices. For two player games, Landsburg \cite{Landsburg} gives a quaternionic computational framework that allows for the identification and classification of the potential Nash equilibria for the EWL quantized game where the intial state is the maximally entangled state $\left|0\right\rangle \otimes \left|0\right\rangle+\left|1\right\rangle \otimes \left|1\right\rangle$. The two classical pure 
strategies available to the players are represented in this quantized game by the $SU(2)$ matrices {\it no flip} ($N$) and {\it flip} ($F$) respectively indicated in (1),
\begin{equation}
\label{eq1}
\left( {{\begin{array}{*{20}c}
 1 \hfill & 0 \hfill \\
 0 \hfill & 1 \hfill \\
\end{array} }} \right),\mbox{ }\left( {{\begin{array}{*{20}c}
 0 \hfill & \eta \hfill \\
 {-\bar {\eta }} \hfill & 0 \hfill \\
\end{array} }} \right)
\end{equation}
where the unit complex number $\eta $ is chosen such that the four outcome states of 
our two player game forms an orthogonal basis of the state 
space $\mathbb{C}P^4$, with standard basis $\left\{ {\left| {00} \right\rangle 
,\left| {01} \right\rangle ,\left| {10} \right\rangle ,\left| {11} 
\right\rangle } \right\}$. These four outcome states are denoted by \textit{NN}, \textit{NF}, 
\textit{FN}, \textit{FF}, where any of these pairs of $N $ and $F$ represent the pure strategy choices of 
the players I and II, respectively. A direct calculation shows for orthogonality of the 
outcome states, by necessity $\eta ^8=1$, so setting 
$$
\eta =\frac{1+i}{\sqrt 2 }
$$
makes the four states \textit{NN}, \textit{NF}, \textit{FN}, \textit{FF} mutually orthogonal. As described above, a pure 
quantum strategy for each player corresponds to a general element of 
$SU(2)$ acting on a player's qubit before the player sends it to the referee, and as described above a classical pure strategy corresponds to the player's choice of one of the 
matrices $N$ or $F$ given above for these actions on their qubits.
%
Exploiting the identification of $SU(2)$ with the unit quaternions, and after identifying each pure 
quantum strategy for each player with a suitable unit quaternion $p$ or $q$, Landsburg shows that 
the probability distribution over the four possible outcomes when the 
players use these strategies is then given by the squares of the 
coefficients of the unit quaternion \textit{pq}. This corresponds in our formalism to a map $L$ from $\prod Q_i$ to $\Delta(ImG)$ as shown in the Figure 3 below. When mixed quantum strategies are played, the map $L$ induces a coordinatization of the map $G^{{\cal{Q}}_{I}}_{*}$ from $\Delta(\prod Q_i)$ to $\Delta(ImG^{{\cal{Q}}_{I}})$. We denote this coordinatized map by $L_{*}$. These maps give Landsburg the computational capability to recognize and classify the potential Nash equilibria of $G^{{\cal{Q}}_{I}}$ and $G^{m{\cal{Q}}_{I}}$. In particular, for zero-sum games $G$ he shows that in the game $G^{m{\cal{Q}}_{I}}$ the players can guarantee to themselves the average of the four possible outcomes of the game $G$ \cite{Landsburg}. Frequently, this is a superior equilibrium payoff than that available to the players restricted to using classical mixed strategies. 

\begin{figure}[h]
\begin{center}
\includegraphics[scale=0.50]{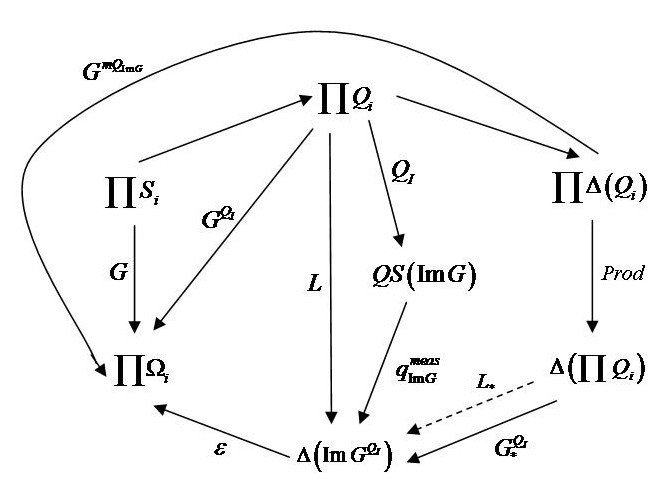}
\caption{Landsburg's maps $L$ and $L_{*}$.}
\end{center}
\end{figure}

We wish to obtain a computational framework similar to Landsburg's in the three player, two pure strategy situation. To this end, we develop an octonionic computational framework that constructs maps analogous to the maps $L$ and $L_{*}$ of Figure 3. Our game will be quantized in a manner similar to that given by Eisert et al \cite{Eisert} and Landsburg \cite{Landsburg} and described above. In our game, the referee initially sends to the three 
players qubits in the maximally entangled state $\left| {000} \right\rangle 
+\left| {111} \right\rangle $. The two classical pure strategies available 
to the players are represented by \textit{no flip} denoted by $N, $and \textit{flip }denoted by $F$, further represented respectively by 
the special unitary matrices in expression (\ref{eq1}), 
where $\eta \in U(\ref{eq1})$ is now chosen so that the eight outcome states of our three 
player game form an orthogonal basis of the state space $\mathbb{C}P^8,$ which 
has standard basis
\begin{equation}
\label{eq2}
\left\{ {\left| {000} \right\rangle ,\left| {001} \right\rangle ,\left| 
{010} \right\rangle ,\left| {011} \right\rangle ,\left| {100} \right\rangle 
,\left| {101} \right\rangle ,\left| {110} \right\rangle ,\left| {111} 
\right\rangle } \right\}
\end{equation}
These eight outcome states are denoted by \textit{NNN}, \textit{NNF}, \textit{NFN}, \textit{NFF}, \textit{FNN}, \textit{FNF}, \textit{FFN}, \textit{FFF}, where any of these 
triples of $N $ and $F$ represent the pure strategy choices of the players I, II and 
III, respectively. Now a direct calculation shows for orthogonality of the 
outcome states, by necessity $\eta ^6=1$, so setting
$$
\eta =\frac{1}{2}+\frac{\sqrt 3 }{2}i
$$
makes the eight states \textit{NNN}, \textit{NNF}, 
\textit{NFN}, \textit{NFF}, \textit{FNN}, \textit{FNF}, \textit{FFN}, \textit{FFF }mutually orthogonal.

As before, the pure quantum strategies for each player are represented by the elements of the Lie group $SU(2)$, which we consider as a copy of the unit quaternions.  We further identify each strategic choice of players I, II, and III with unit octonions 
$s$, $t$, and $u$ respectively, where each of $s$, $t$, and $u$ lies in a particular, possibly different, 
copy of the unit quaternions embedded in the octonions. The probability 
distribution over our eight possible outcomes is then shown to be determined 
by an expression involving the associated triple product $(st)u$ of the 
octonions $s$, $t$, and $u$. The associated nature of this product is in fact 
natural as the octonions are in general non-associative. As in Landsburg's 
case, these identifications and the resulting computational efficiency allows us to examine the effect on the payoffs to each player of 
the game when players use mixtures, superpositions, or even mixed superpositions 
of their classical pure strategies.

\section{Octonions}

\begin{figure}
\begin{center}
\includegraphics[scale=0.80]{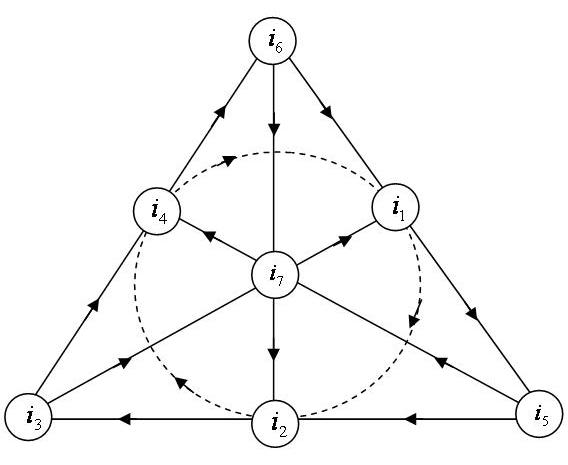}
\caption{An edge oriented Fano plane.}
\end{center}
\end{figure}

An excellent reference on the mathematical structure of the quaternions and 
octonions is \cite{Conway}. The octonions $\mathbb{O}$ are a non-associative 
8-dimensional, normed, real division algebra with elements given as
\begin{equation}
\label{eq3}
a_0 +\sum\limits_{j=1}^7 {a_j } i_j 
\end{equation}
where $a_0$ and the $a_j$ are elements of the real numbers $\mathbb{R}$, and the $i_j $'s have the property that $i_j 
^2=-1$. Moreover, these seven $i_j $'s (along with the real number 1) can be used to 
form seven copies of the quaternions $\mathbb{H}$ embedded within $\mathbb{O}$ 
as certain triple products of the $i_1 \ldots i_7 $ are also equal to $-1$. A 
mnemonic for which of these triple products are in fact equal to $-1$ is given 
by an edge oriented Fano plane illustrated in Figure 4. This Fano plane 
indicates how certain octonionic triple products work. In particular, when the $i_j ,i_k 
,i_l $ are cyclically ordered as in the lines of the edge oriented Fano plane of Figure 4, then $i_j ^2=i_k ^2=i_l ^2=i_j i_k i_l =-1$. This shows that in general $i_j 
i_k =i_l =-i_k i_j $. 

We require three copies of the quaternions $\mathbb{H}$ embedded in the 
octonions $\mathbb{O}$ possessing a common embedded copy of the complex 
numbers $\mathbb{C}$. We choose copies $\mathbb{H}_1$, $\mathbb{H}_2$, $\mathbb{H}_3$ of the quaternions $\mathbb{H}$ with octonionic bases 
$$
\left\{1,i_1 ,i_2 ,i_4 \right\}, \left\{1,i_1 ,i_5 ,i_6 \right\}, \left\{ 1,i_1 ,i_3 
,i_7 \right\}
$$
respectively. Note $\mathbb{H}_1$, $\mathbb{H}_2$, $\mathbb{H}_3$ have a common embedded copy of the complex 
numbers $\mathbb{C}$ with octonionic basis $\left\{ {1,i_1 } \right\}$. We focus our 
attention on the unit $S^3$'s in each of these four dimensional copies of 
$\mathbb{H}$ and consider each such $S^3$ as a ``longitude'' of the unit 
octonions which as a set forms a seven dimensional sphere $S^7 \subset 
\mathbb{O}$.

We now identify the pure quantum strategies available to each player with a 
particular unit octonion, considering the complex numbers involved as 
elements of our common embedded copy of $\mathbb{C}$. As usual, these quantum 
strategies each correspond to an element of $SU(2)$. Recall that elements of the group $SU(2)$ are $2\times 2$ unitary complex matrices 
with determinant one, and can be written in the form
\begin{equation}
\label{eq4}
\left( {{\begin{array}{*{20}c}
 x \hfill & y \hfill \\
 {-\bar {y}} \hfill & {\bar {x}} \hfill \\
\end{array} }} \right)
\end{equation}
If player 1 chooses his pure quantum strategy corresponding to the $SU(2)$ matrix
\begin{equation}
\label{eq5}
\left( {{\begin{array}{*{20}c}
 A \hfill & B \hfill \\
 {-\bar {B}} \hfill & {\bar {A}} \hfill \\
\end{array} }} \right)
\end{equation}
where $A=a_0 +a_1 i$, $B=b_0 +b_1 i$, identify this strategy with 
the unit octonion given by 
\begin{equation}
\label{eq6}
\begin{array}{l}
s_{00} =A+B\bar{\eta }i_4 = a_0 +a_1 i_1 +\left( {b_0 +b_1 i_1 } 
\right)\left( {\frac{1}{2}-\frac{\sqrt 3 }{2}i_1 } \right)i_4 \\ 
= a_0 +a_1 i_1 +\left( {\frac{\sqrt 3 }{2}b_0 -\frac{1}{2}b_1 } \right)i_2 
+\left( {\frac{1}{2}b_0 +\frac{\sqrt 3 }{2}b_1 } \right)i_4 \\ 
 \end{array}
\end{equation}
The subscript 00 on $s$ is used to track various sign changes on two of the
coefficients in the expression for $s$, namely, $a_0$ and $a_1$. A positive sign will be represented by 0 and a sign change to 
a negative in the expression for $s $ will be represented by 1. This notation 
will be used below to indicate the appropriate coefficient in the resulting 
probability distribution in our EWL quantized version of $G$.

Similarly, if player 2 chooses a quantum strategy corresponding to the $SU(2)$ matrix
\begin{equation}
\label{eq7}
\left( {{\begin{array}{*{20}c}
 P \hfill & Q \hfill \\
 {-\bar {P}} \hfill & {\bar {Q}} \hfill \\
\end{array} }} \right)
\end{equation}
where $P=p_0 +p_1 i$, $Q = q_0 +q_1 i$, identify this strategy with the unit octonion given by
\begin{equation}
\label{eq8}
\begin{array}{l}
 t_{00} =P+Q\bar {\eta }i_6 \\ 
 =p_0 +p_1 i_1 +\left( {\frac{\sqrt 3 }{2}q_0 -\frac{1}{2}q_1 } \right)i_5 
+\left( {\frac{1}{2}q_0 +\frac{\sqrt 3 }{2}q_1 } \right)i_6 \\ 
 \end{array}
\end{equation}
And if player 3 chooses the quantum strategy corresponding to the $SU(2)$ matrix
\begin{equation}
\label{eq9}
\left( {{\begin{array}{*{20}c}
 E \hfill & F \hfill \\
 {-\bar {F}} \hfill & {\bar {E}} \hfill \\
\end{array} }} \right)
\end{equation}
where $E=e_0 +e_1 i$, $F=f_0 +f_1 i$, identify this strategy with the unit octonion given by
\begin{equation}
\label{eq10}
\begin{array}{l}
 u_{00} =E+F\bar {\eta }i_7 \\ 
 =e_0 +e_1 i_1 +\left( {\frac{\sqrt 3 }{2}f_0 -\frac{1}{2}f_1 } \right)i_3 
+\left( {\frac{1}{2}f_0 +\frac{\sqrt 3 }{2}f_1 } \right)i_7 \\ 
 \end{array}
\end{equation}
with the subscripts for $t$ and $u$ behaving as they do for $s$, with the subscripts on $t$ refering to sign changes for $p_0$ and $p_1$ and the subscripts for $u$ refering to sign changes on $e_0$ and $e_1$. We are now ready to describe our analogue of Landsburg's map $L$. 

\begin{mythm}
If in the maximally entangled three player, two strategy EWL quantized game, 
player 1 plays the pure quantum strategy given in (\ref{eq5}), player 2 the pure 
quantum strategy given in (\ref{eq7}), and player 3 the pure quantum strategy given 
in (\ref{eq9}), then the probability distribution over the outcomes \textit{NNN}, \textit{NNF}, \textit{NFN}, \textit{NFF}, \textit{FNN}, \textit{FNF}, 
\textit{FFN}, \textit{FFF} is given by
$$
\begin{array}{l}
 pr(NNN)=\left[ {\pi _0 \left( {\frac{\left( {s_{10} t_{10} } 
\right)u_{01} +\left( {s_{01} t_{10} } \right)u_{01} }{2}} \right)} 
\right]^2+\left[ {\pi _0 \left( {\frac{\left( {s_{10} t_{10} } 
\right)u_{01} -\left( {s_{01} t_{10} } \right)u_{01} }{2}} \right)} 
\right]^2 \\ 
 pr(FFF)=\left[ {\pi _1 \left( {\frac{\left( {s_{10} t_{10} } 
\right)u_{01} +\left( {s_{01} t_{10} } \right)u_{01} }{2}} \right)} 
\right]^2+\left[ {\pi _1 \left( {\frac{\left( {s_{10} t_{10} } 
\right)u_{01} -\left( {s_{01} t_{10} } \right)u_{01} }{2}} \right)} 
\right]^2 \\ 
 pr(NFF)=\left[ {\pi _2 \left( {\frac{\left( {s_{01} t_{00} } 
\right)u_{00} +\left( {s_{10} t_{00} } \right)u_{00} }{2}} \right)} 
\right]^2+\left[ {\pi _2 \left( {\frac{\left( {s_{01} t_{00} } 
\right)u_{00} -\left( {s_{10} t_{00} } \right)u_{00} }{2}} \right)} 
\right]^2 \\ 
 pr(FNN)=\left[ {\pi _4 \left( {\frac{\left( {s_{01} t_{00} } 
\right)u_{00} +\left( {s_{10} t_{00} } \right)u_{00} }{2}} \right)} 
\right]^2+\left[ {\pi _4 \left( {\frac{\left( {s_{01} t_{00} } 
\right)u_{00} -\left( {s_{10} t_{00} } \right)u_{00} }{2}} \right)} 
\right]^2 \\ 
 pr(FNF)=\left[ {\pi _5 \left( {\frac{\left( {s_{01} t_{00} } 
\right)u_{00} +\left( {s_{10} t_{00} } \right)u_{00} }{2}} \right)} 
\right]^2+\left[ {\pi _5 \left( {\frac{\left( {s_{01} t_{00} } 
\right)u_{00} -\left( {s_{10} t_{00} } \right)u_{00} }{2}} \right)} 
\right]^2 \\ 
 pr(NFN)=\left[ {\pi _6 \left( {\frac{\left( {s_{01} t_{00} } 
\right)u_{00} +\left( {s_{10} t_{00} } \right)u_{00} }{2}} \right)} 
\right]^2+\left[ {\pi _6 \left( {\frac{\left( {s_{01} t_{00} } 
\right)u_{00} -\left( {s_{10} t_{00} } \right)u_{00} }{2}} \right)} 
\right]^2 \\ 
 pr(FFN)=\left[ {\pi _3 \left( {\frac{\left( {s_{10} t_{10} } 
\right)u_{01} +\left( {s_{01} t_{10} } \right)u_{01} }{2}} \right)} 
\right]^2+\left[ {\pi _3 \left( {\frac{\left( {s_{10} t_{10} } 
\right)u_{01} -\left( {s_{01} t_{10} } \right)u_{01} }{2}} \right)} 
\right]^2 \\ 
 pr(NNF)=\left[ {\pi _7 \left( {\frac{\left( {s_{10} t_{10} } 
\right)u_{01} +\left( {s_{01} t_{10} } \right)u_{01} }{2}} \right)} 
\right]^2+\left[ {\pi _7 \left( {\frac{\left( {s_{10} t_{10} } 
\right)u_{01} -\left( {s_{01} t_{10} } \right)u_{01} }{2}} \right)} 
\right]^2 \\ 
 \end{array}
 $$
where $\pi _j $ is the projection function projecting the input 
octonion onto the subspace of $\mathbb{O}$ spanned by the vector $i_j $ considering the real number 1 as $i_0$.
\end{mythm}

The proof appears in \cite{ABK}. While slightly more complicated than Landsburg's quaternionization for two player games, the $L$ map given in Theorem 1 and the analogous $L_{*}$ map are robust enough to demonstrate the existence of a mixed quantum Nash Equilibrium in $G^{m{\cal{Q}}_I}$ in which each player employs the uniform distribution over their pure quantum strategies \cite{ABK}. In this equilibrium each player receives the average of their 8 possible classical payouts.  This is enough to completely change the behavior of the Nash-Shapley Poker Model \cite{Nash}, see \cite{Bleiler2}; and gives new equilibria for the mixed quantum versions of certain three player Dilemma Games, see \cite{Ahmed}.

\end{document}